\documentclass[twocolumn,amssymb,amsmath,aps,pra,groupedaddress,nobibnotes,showpacs]{revtex4}
\usepackage{graphicx}

\begin{document}

\title{Observation of correlated-photon statistics using a single detector}

\author{Yoon-Ho Kim}\email{kimy@ornl.gov}
\affiliation{Center for Engineering Science Advanced Research, Computer Science and Mathematics Division\\ Oak Ridge National Laboratory, Oak Ridge, Tennessee 37831}

\author{Warren P. Grice}
\affiliation{Center for Engineering Science Advanced Research, Computer Science and Mathematics Division\\ Oak Ridge National Laboratory, Oak Ridge, Tennessee 37831}

\date{\today}

\begin{abstract}
We report experimental observations of correlated-photon statistics in the single-photon detection rate. The usual quantum interference in a two-photon polarization interferometer always accompanies a dip in the single detector counting rate, regardless of whether a dip or peak is seen in the coincidence rate. This effect is explained by taking into account all possible photon number states that reach the detector, rather than considering just the state post-selected by the coincidence measurement. We also report an interferometeric scheme in which the interference peak or dip in coincidence corresponds directly to a peak or dip in the single-photon detection rate.
\end{abstract}


\maketitle

In interference experiments involving two-photon fields of spontaneous parametric down-conversion (SPDC), quantum interference effects are typically observed in the rate of coincidence counts between two detectors, while the single-detector count rate is expected to be featurelessly constant \cite{single}.  (A good example is the two-photon anti-correlation dip-peak experiment  \cite{bell,hom,type2,ultrafast}.) Indeed, this would be the case if the single-photon detectors available today were truly 100\% efficient and were able to resolve multi-photon excitations. However, all commercially available solid-state single-photon detectors today rely on the avalanche proccess of Si or InGaAs/InP photodiodes. Therefore, even with 100\% efficiency, these detectors cannot resolve photon number. This effect usually does not reveal any information about the incident state, as it simply reduces the overall detection efficiency. 

In certain cases, however, the single-detector count rate does provide information about the incident state. This was first demonstrated in Ref.~\cite{steinberg}, where a quantum interference effect in a two-photon interferometer was employed to change the photon statistics at a single detector. It was found that the coincidence dip associated with the photon bunching effect at a beamsplitter was accompanied by a dip in the single detector counting rate, as well. At the center of the coincidence dip, the photons always leave the interferometer (or the beamsplitter) together. Thus, a detector monitoring one of the output ports of the interferometer ``sees'' either $|0\rangle$ or $|2\rangle$, but never $|1\rangle$. Compared to the photon statistics outside the coincidence dip, where the two photons are randomly distributed to the detectors, a single detector sees fewer photon events in the coincidence dip, even though the mean photon number does not change. Because the detector is unable to distinguish between $|1\rangle$ and $|2\rangle$, a single-detector dip is observed.

In this paper, we first confirm the dip effect in the single-detector count rate using a different experimental setup. We also measure the single-detector count rate with the interferometer designed for a  coincidence peak, rather than a dip. Somewhat surprisingly, the coincidence peak is not reflected as a peak in the the single-detector count rate. Instead, the singles rate reveals a dip, just as if the interferometer were aligned for a coincidence dip. This result can be explained by taking into account all possible photon number states that reach the detector, rather than just the state post-selected by the coincidence measurement. Finally, we present an experiment in which the coincidence peak or dip directly corresonds to a dip or peak in the singles rate. 

\begin{figure}[t]
\includegraphics[width=2.8in]{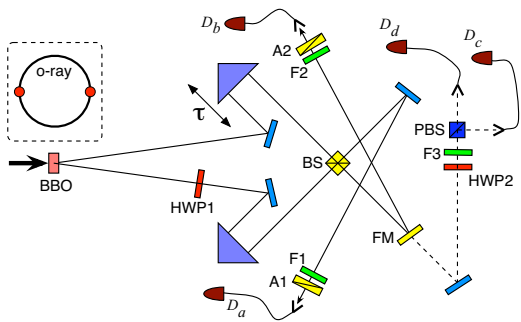}
\caption{\label{fig:setup1}Outline of experimental setup. FM is a flipper mirror, A1 and A2 are polarizers, and F1 $\sim$ F3 are spectral filters.}
\end{figure}

Consider the experimental setup shown in Fig.~\ref{fig:setup1}. SPDC photon pairs are generated in a 2 mm thick type-I BBO crystal pumped with a 351.1 nm argon ion laser. The FWHM of spectral filters F1 and F2 were 3 nm and the coincidence window for all measurements was about 3 nsec. The non-collinear 702.2 nm signal and idler photons are brought together on a beamsplitter and one arm of the interferometer can be adjusted by a computer-controlled DC motor. The non-collinear arrangement avoids the problematic second-order (of the field) interference effect reported in Ref.~\cite{steinberg}. 

\begin{figure}[t]
\includegraphics[width=3in]{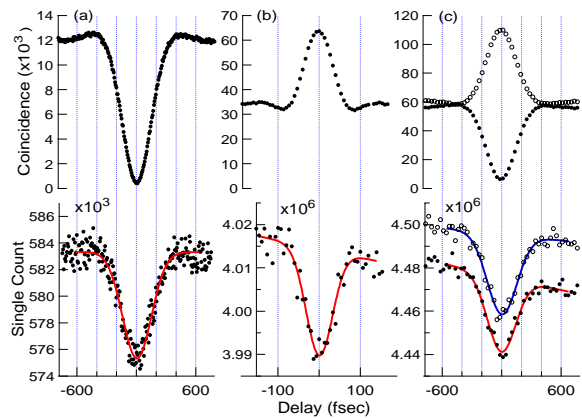}
\caption{\label{fig:data1}Experimental data. (a) $D_a$-$D_b$ coincidence dip. 40 seconds each point. (b) $D_c$-$D_d$ coincidence. 10 seconds  each point. (c) Polarization correlation measurement with $D_a$ and $D_b$. Coincidence peak (dip) is measured for polarizer angles A1/A2 = $45^{\circ}/-45^{\circ}$ (= $45^{\circ}/45^{\circ}$). 40 seconds each point. }
\end{figure}

With HWP1, A1, and A2 removed from the apparatus, the usual coincidence dip is obtained by scanning the delay $\tau$ \cite{hom}. The experimental data for this measurement is shown in Fig.~\ref{fig:data1}(a). Note that both the coincidence rate and the single-detector rate show dips as the delay is scanned. Note, also, that the two dips have the same widths. The dip in the single-count rate can be understood more formally as follows. If $\eta$ is the single-photon detection efficiency, then the probability of a detection event in the presence of two photons is given by $\eta + (1-\eta)\eta=2\eta-\eta^{2}$ \cite{steinberg}. The overall single-detector counting rate can then be written as 
\begin{equation}
R \propto P_{1}\eta +P_{2}(2\eta-\eta^{2}),\label{eq}
\end{equation} 
where $P_{1}$ and $P_{2}$ are the probabilities that one and two photons, respectively, are incident on the detector.

The photon statistics at the output ports of the beamsplitter are determined entirely by the delay $\tau$ in this case. If $\tau>\tau_{c}$, where $\tau_{c}$ is the coherence time of the single-photon wavepacket, incident photons simply scatter independently, resulting in four possible events at the output: (i) both photons reflected, (ii) both photons transmitted, (iii) both photons end up at $D_a$, and (iv) both photons end up at $D_b$. Since each of these events is equally likely, the probabilities that a particular output port, $D_a$ or $D_b$, contains zero, one, and two photons are $P_{b0}=1/4$, $P_{b1}=1/2$, and $P_{b2}=1/4$. If, on the other hand, $\tau=0$, quantum interference causes amplitudes for (i) and (ii) to sum to zero \cite{bell,hom,type2,ultrafast}. In this case, $P_{b0}=1/2$, $P_{b1}=0$, and $P_{b2}=1/2$. With these probabilities, which are summarized in Table \ref{table},  Eq.(\ref{eq}) yields the single-detector counting rates
\begin{equation}
R(\tau>\tau_c) \propto \eta - \frac{1}{4}\eta^2, \hspace{0.5cm}
R(\tau=0) \propto \eta - \frac{1}{2}\eta^2.
\end{equation}
The above result clearly shows that a dip in the singles rate is expected to accompany a dip in the coincidence rate between detectors $D_a$ and $D_b$.

The coincidence dip in this case can be regarded as the signature of the state $\frac{1}{\sqrt{2}}(|2,0\rangle+|0,2\rangle)$ exiting the beamsplitter. When $\tau=0$, each detector receives either zero photons or two photons, but never one photon. Consider, now, the case in which a peak is observed in the coincidence rate. This is accomplished in our setup by removing the flipper mirror, thus directing one output of the beamsplitter to detectors $D_c$ and $D_d$. The detectors are preceded by a halfwave plate and polarization beamspitter, which act together as a 50/50 beamsplitter. The FWHM of the spectral filter F3 was 20 nm. When $\tau=0$, the path exiting the beamsplitter BS contains either zero or two photons, since this delay corresponds to the center of the coincidence dip for detectors $D_a$ and $D_b$. With a higher probability of finding two photons in the exit path (1/2 for $\tau=0$ vs. 1/4 for $\tau>\tau_{c}$), a coincidence peak is observed between $D_c$ and $D_d$, as shown in Fig.~\ref{fig:data1}(b)  \cite{rarity}.

It is tempting to regard such a peak as signalling the presence of the state $|1,1\rangle$. If this were true, then a peak in the single-detector counting rate would also be expected, since every photon pair emission would lead to exactly one photon at each detector. However, this is not the case. Instead of a peak in the single-photon counting rate, a dip is observed just as in the case of the coincidence dip between $D_a$ and $D_b$. This rather unexpected result can be explained by considering conditional probabilities at the second beamsplitter. The probabilities that zero, one, and two photons are incident on, for example, detector $D_c$ are
\begin{eqnarray}
P_{0}&=&P_{b0}P_{00}+P_{b1}P_{10}+P_{b2}P_{20},\nonumber\\
P_{1}&=&P_{b0}P_{02}+P_{b1}P_{11}+P_{b2}P_{21}=P_{b1}P_{11}+P_{b2}P_{21},\nonumber\\
P_{2}&=&P_{b0}P_{02}+P_{b1}P_{12}+P_{b2}P_{22}=P_{b2}P_{22},\label{prob}
\end{eqnarray}
where, as defined above, $P_{b0}$, $P_{b1}$, and $P_{b2}$ are the probabilities that zero, one, and two photons leave the first beamsplitter, respectively. The conditional probabilities $P_{ij}$ are defined as the probabilities that $j$ photons will exit port $c$ of the second beamsplitter, given $i$ incident photons. These conditional probablities are independent of the delay $\tau$ and are summarized in Table \ref{table}. With these quantities, Eq.~(\ref{eq}) yields
\begin{equation}
R(\tau>\tau_c) \propto \frac{1}{2}\eta - \frac{1}{16}\eta^2, \hspace{0.5cm}
R(\tau=0) \propto \frac{1}{2}\eta - \frac{1}{8}\eta^2.\label{eq:dip2}
\end{equation}
Here, we clearly see that a dip in the single-detector counting rate should occur even in this case. Thus, while a coincidence detection signals one photon in each output port of the second beamsplitter, it should not be assumed that the output state is $|1,1\rangle$. In this case, there are clearly instances in which the two photons exit the second beamsplitter (HWP2-PBS set) via the same port.

\begingroup
\squeezetable
\begin{table*}
\caption{\label{table} Summary of probabilities that a particular output port contains zero, one, and two photons for the three distinct experimental conditions considered in this paper. BG refers to the background random probabilities which occurs when $\tau>\tau_c$.}
\begin{ruledtabular}
\begin{tabular}{ccccccccccccc}
\multicolumn{5}{c}{Two photons have the same polarization} & \multicolumn{5}{c}{Two photons are orthogonally polarized} & \multicolumn{3}{c}{Deterministic case (Fig.~\ref{fig:setup2})} \\
\multicolumn{2}{c}{At Beamsplitter} & \multicolumn{3}{c}{Cond. prob. at $D_c$ ($D_d$)} & At BS & \multicolumn{4}{c}{Prob. at $D_a$ ($D_b$) with $\pm45^\circ$ polarizer} & BG & Dip & Peak \\
$\tau>\tau_c$ & $\tau=0$ & \multicolumn{3}{c}{Prob. independent of $\tau$} & \multicolumn{3}{c}{Prob. independent of $\tau$} & $\tau>\tau_c$ & $\tau=0$ & $\tau>\tau_c$ & $\tau=0$ & $\tau=0$ \\ 
$P_{b0}=\frac{1}{4}$ & $P_{b0}=\frac{1}{2}$ & $P_{00}=1$ & $P_{10}=\frac{1}{2}$ & $P_{20}=\frac{1}{4}$ & $P_{b0}=\frac{1}{4}$ & $P_{00}=1$ & $P_{10}=\frac{1}{2}$ & $P_{20}=\frac{1}{4}$ & $P_{20}=\frac{1}{2}$ & $P_0=\frac{1}{4}$ & $P_0=\frac{1}{2}$ & $P_0=0$ \\
$P_{b1}=\frac{1}{2}$ & $P_{b1}=0$ & $P_{01}=0$ & $P_{11}=\frac{1}{2}$ & $P_{21}=\frac{1}{2}$ & $P_{b1}=\frac{1}{2}$ & $P_{01}=0$ & $P_{11}=\frac{1}{2}$ & $P_{21}=\frac{1}{2}$ & $P_{21}=0$ & $P_1=\frac{1}{2}$ & $P_1=0$ & $P_1=1$\\
$P_{b2}=\frac{1}{4}$ & $P_{b2}=\frac{1}{2}$ & $P_{02}=0$ & $P_{12}=0$ & $P_{22}=\frac{1}{4}$ & $P_{b2}=\frac{1}{4}$ & $P_{02}=0$ & $P_{12}=0$ & $P_{22}=\frac{1}{4}$ & $P_{22}=\frac{1}{2}$ & $P_2=\frac{1}{4}$ & $P_2=\frac{1}{2}$ & $P_2=0$
\end{tabular}
\end{ruledtabular}
\end{table*}
\endgroup

Let us now consider the case in which the coincidence peak-dip may be observed in a single apparatus: HWP1 rotates the photon polarization by $90^\circ$ and polarizers are inserted in front of the detectors $D_a$ and $D_b$. [This is a typical Bell-experiment setup.] When $\tau=0$, polarizer settings of $A1/A2=45^\circ/45^\circ$ result in a null in the coincidence rate, while settings of $A1/A2=45^\circ/-45^\circ$ result in a coincidence peak \cite{bell,type2,ultrafast}. The experimental data for these measurements are shown in Fig.~\ref{fig:data1}(c). The coincidence measurements show the expected peak and dip, while the single-count measurements, once again, yield  dips in both cases. 

As before, these results can be understood by taking into account all possible photon number states at the detector, rather than just the states post-selected by the coincidence measurement. Since the two input photons are orthogonally polarized, they exit the beamsplitter BS independently, regardless of the delay $\tau$.  Therefore, $P_{b0}=1/4, P_{b1}=1/2$, and $P_{b2}=1/4$ in both modes $a$ and $b$ before the polarizers. At the polarizer ($\pm45^\circ$ oriented), single photons are passed only half the time, regardless of the delay. When two photons are incident, however, the result depends on the delay $\tau$. The orthogonally polarized photons scatter randomly  for $\tau>\tau_{c}$, while quantum interference occurs when $\tau=0$. In the latter case, the two photons are either both blocked or both passed at the polarizer. With these probabilities, which are summarized in Table I, Eqs.(\ref{eq}) and (\ref{prob}) yield  the same overall single-detector counting rates as given in Eq.~(\ref{eq:dip2}), which predict a dip in the single-detector rate, regardless of whether the coincidence shows a peak or a dip. 

As in the previous case, the presence of a coincidence peak does not indicate the state $|1,1\rangle$ exiting the beamsplitter. Indeed, in the Bell-state generation scheme, the orthogonally polarized photons always exit the beamsplitter in random fashion. When the photons exit the beamsplitter via different ports and a coincidence is registered with orthogonally oriented polarizers (polarizer settings for a coincidence peak), it is certainly the case that one photon reached each detector. Because of the polarization entanglement between the two photons, the rate at which coincidences are registered is higher when $\tau=0$. It is not the case, however, that the photons always exit via different ports. These other cases, in which the photons exit the beamsplitter together, do not lead to coincidences, but they do contribute to the singles rates. Therefore, the complete description of the state reaching the detectors must include not only the $|1,1\rangle$ term, but also the terms which lead to photons at only one detector. 

It should also be pointed out that, in contrast to the case in which the photons have the same polarizations when they reach the beamsplitter [this setup leads to the experimental data shown in Fig.~\ref{fig:data1}(a)], the presence of a coincidence dip in a Bell-sate generation scheme does not indicate the state $\frac{1}{\sqrt{2}}(|2,0\rangle+|0,2\rangle)$. The state reaching the detectors must also include the terms $|1,0\rangle$ and $|0,1\rangle$. These terms are present because the polarization entanglement ensures that, for the cases in which the photons exit the beamsplitter via different ports toward identically oriented polarizers (polarizer settings for a coincidence dip), only one of the two photons will reach the detectors.

It is also interesting to note that the dip in the singles rate is due to a quantum interference effect that differs from the effect leading to the interference features in the coincidence rate. In the latter case, coincidence detection collapses the two-photon state to a polarization-entangled state (the terms $|2,0\rangle$ and $|0,2\rangle$ do not lead to coincidences). The coincidence rate for this entangled state depends on the (relative) orientations of the two polarizers. The interference observed in the singles rate is different not only because only a single polarizer is required, but also because the terms discarded in coincidence detection become important. The singles rate is independent of $\tau$ when single photons reach the polarizer, but when two photons are present, photon bunching occurs when $\tau=0$, i.e., the photons are passed or blocked as a pair at the $\pm45^\circ$ polarizer.

\begin{figure}[t]
\includegraphics[width=2.8in]{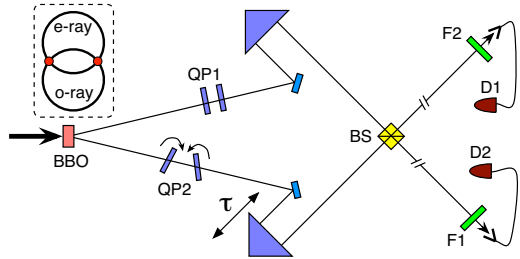}
\caption{\label{fig:setup2}Outline of experimental setup.}
\end{figure}

An obvious drawback to the Bell-state generation scheme is that it is not possible to deterministically generate (or switch between) the states $\frac{1}{\sqrt{2}}(|2,0\rangle+|0,2\rangle)$ and $|1,1\rangle$. If it were possible to generate these states without relying on post-selective measurements, then photon pairs with well-known quantum states would be available for further processing or for use in other applications. Unlike the schemes discussed thus far, such a method would be characterized by single-detector counting rates that would differ for the coincidence peak and dip. That is, the state $\frac{1}{\sqrt{2}}(|2,0\rangle+|0,2\rangle)$, which would yield no coincidences, would lead to probablities $P_0=1/2$, $P_1=0$, and $P_2=1/2$ for a single detector. Meanwhile, the state $|1,1\rangle$ would yield only coincidences and would lead to single-detector probabilities of $P_0=0$, $P_1=1$, and $P_2=0$. According to Eq.~(\ref{eq}), the single-detector counting rates would be 
\begin{equation}
R_{peak}(\tau=0) \propto \eta,\hspace{0.5cm} R_{dip}(\tau=0) \propto \eta -\frac{1}{2}\eta^2,\label{eq4}
\end{equation}
for these two cases. Thus, the singles rate would mirror the coincidence rate, i.e., it would increase (decrease) in the presence of a coincidence peak (dip).

\begin{figure}[t]
\includegraphics[width=2.8in]{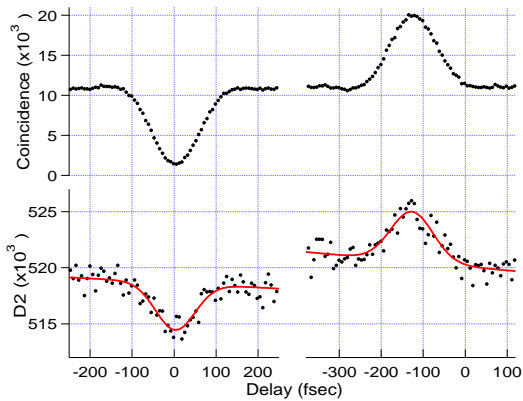}
\caption{\label{fig:data2}Experimental data. Data accumulation time is 10 sec. The coincidence peak-dip visibility is about 87\%.}
\end{figure}

Fig.~\ref{fig:setup2} shows the outline of the apparatus used to generate the above mentioned two-photon number states. A 3 mm thick type-II BBO crystal is pumped by an ultrafast pulse with central wavelength of 390 nm and pulse durations of approximately 120 fsec. Pairs of photons with center wavelengths of 780 nm emerge from the crystal into two separate cones, one belonging to the e-ray (V-polarized) and the other belonging to the o-ray (H-polarized) of the crystal. Here, we are interested in the photons emitted into the intersections of the two cones. These two spatial modes make up the two input ports of an ordinary beamsplitter. The FWHM of the spectral filters F1 and F2 was 20 nm. With the interferometer properly balanced, it is possible to switch between the two states $|1,1\rangle$ and $\frac{1}{\sqrt{2}}(|2,0\rangle+|0,2\rangle)$ simply by tilting the quartz plates QP2. Detailed discussions of the interferometer can be found elsewhere \cite{kimgrice,mattle}. 

The experimental results are shown in Fig.~\ref{fig:data2}. With QP2 normal to the beam path, a coincidence peak was observed, while an orientation of approximately $23.5^\circ$ produced a coincidence dip. Unlike the experiments described earlier, the coincidence features in this experiment are reflected in the single-detector counting rates, shown in the lower portion of Fig.~\ref{fig:data2}. This suggests that all the photons reaching the detectors are either in the state $\frac{1}{\sqrt{2}}(|2,0\rangle+|0,2\rangle)$ or in the state $|1,1\rangle$, depending on the phase setting of QP2. 

In summary, we have reported the experimental observation of various photon statistics observed in single-photon detection rates in different quantum interferometric schemes. The observed dip in the single-detector counting rate is the combined result of quantum interference and the inability of the detectors to distinguish two-photon excitations from  single-photon excitations. In addition, we showed that two-photon number states prepared in a typical two-photon interferometer are post-selective. As a result, a dip in the single detector counting rate was observed regardless of whether a dip or peak was seen in the coincidence rate in a typical two-photon interferometer. We concluded with an interference experiment in which two-photon number states can be prepared in a deterministic fashion. This was confirmed by observing a correspondence in the peak and dip in single-detector counting rates with the peak and dip in coincidence rates. 

This research was supported in part by the U.S. DOE, Office of Basic Energy Sciences, the National Security Agency, and the LDRD Program of the Oak Ridge National Laboratory, managed for the U.S. DOE by UT-Battelle, LLC, under contract No.~DE-AC05-00OR22725.


\begin{thebibliography}{}

\bibitem{single} Although not very common, quantum interference can also be observed in the single count rate. See X.Y. Zou, L.J. Wang, and L. Mandel, Phys. Rev. Lett. \textbf{67}, 318 (1991); Y.-H. Kim \textit{et al.}, Phys. Rev. A \textbf{61}, 051803(R), (2000).

\bibitem{bell} Y.H. Shih and C.O. Alley in \textit{Proc. 2nd Int. Symp. Found. Quantum Mechanics}, ed. M. Namiki \textit{et al}. (Physical Society of Japan, Tokyo, 1987); Y.H. Shih and C.O. Alley, Phys. Rev. Lett. \textbf{61}, 2921 (1988); Z.Y. Ou and L. Mandel, \textit{ibid}. \textbf{61}, 50 (1988); P.G. Kwiat, A.M. Steinberg, and R.Y. Chiao, Phys. Rev. A \textbf{45}, 7729 (1992).

\bibitem{hom} C.K. Hong, Z.Y. Ou, and L. Mandel, Phys. Rev. Lett. \textbf{59}, 2044 (1987).

\bibitem{type2} Y.H. Shih and A.V. Sergienko, Phys. Lett. A \textbf{186}, 29 (1994); \textit{ibid.} \textbf{191}, 201 (1994); T.B. Pittman \textit{et al.}, Phys. Rev. Lett. \textbf{77}, 1917 (1996).

\bibitem{ultrafast} G. Di Giuseppe \textit{et al.}, Phys. Rev. A \textbf{56}, R21 (1997); W.P. Grice \textit{et al.}, Phys. Rev. A \textbf{57}, R2289, (1998); Y.-H. Kim \textit{et al.}, Phys. Rev. A \textbf{64}, 011801 (2001).

\bibitem{steinberg} K.J. Resch, J.S. Lundeen, and A.M. Steinberg, Phys. Rev. A \textbf{63}, 020102 (2001). 

\bibitem{rarity} J.G. Rarity and P.R. Tapster, J. Opt. Soc. Am. B \textbf{6}, 1221 (1989).

\bibitem{kimgrice} Y.-H. Kim and W.P. Grice, quant-ph/0304086.

\bibitem{mattle} K. Mattle \textit{et al.}, Phys. Rev. Lett. \textbf{76}, 4656 (1996)

\end{thebibliography}
\end{document}